\documentclass[%
 aip,
 amsmath,amssymb,
 reprint,%
]{revtex4-1}

\usepackage{dcolumn,amsmath}
\usepackage{graphicx}
\usepackage{bm}
\usepackage{hyperref}

\usepackage[utf8]{inputenc}
\usepackage[T1]{fontenc}

\begin{document}

\title{Approaching meV level for transition energies in the radium monofluoride molecule RaF and radium cation Ra$^+$ by including quantum-electrodynamics effects}

\date{09.04.2021}

\begin{abstract}
Highly accurate theoretical predictions of transition energies in the radium monofluoride molecule, $^{226}$RaF and radium cation, $^{226}$Ra$^+$, are reported. The considered transition $X~^2\Sigma_{1/2} \to A~^2\Pi_{1/2}$ in RaF is one of the main features of this molecule and can be used to laser cool RaF for subsequent measurement of the electron electric dipole moment. For molecular and atomic predictions we go beyond the Dirac-Coulomb Hamiltonian and treat high-order electron correlation effects within the coupled cluster theory with the inclusion of quadruple and ever higher amplitudes. Effects of quantum electrodynamics (QED) are included non-perturbatively using the model QED operator that is implemented now for molecules. It is shown that the inclusion of QED effects in molecular and atomic calculations is a key ingredient in resolving the discrepancy between the theoretical values obtained within the Dirac-Coulomb-Breit Hamiltonian and the experiment. The remaining deviation from the experimental values is within a few meV. This is more than an order of magnitude better than the ``chemical accuracy'', 1 kcal/mol=43 meV, that is usually considered as a guiding thread in theoretical molecular physics.
\end{abstract}

\author{Leonid V.\ Skripnikov}
\email{skripnikov\_lv@pnpi.nrcki.ru,\\ leonidos239@gmail.com}
\homepage{http://www.qchem.pnpi.spb.ru}
\affiliation{Petersburg Nuclear Physics Institute named by B.P. Konstantinov of National Research Centre
``Kurchatov Institute'', Gatchina, Leningrad District 188300, Russia}
\affiliation{Saint Petersburg State University, 7/9 Universitetskaya nab., St. Petersburg, 199034 Russia}

\maketitle

\section{Introduction}
Experiments with highly charged ions, neutral atoms and molecules provide information about the fundamental interactions, allow testing the Standard model's predictions and searching for New physics~\cite{Safronova:18}. The accuracy of theoretical predictions of properties of highly charged ions with a few electrons is extremely high and allows one to extract fundamental constants. Experiments with heavy neutral atoms allow studying parity nonconserving interactions~\cite{Wood:1997}. 
The accuracy of theoretical predictions for highly charged ions is provided by the use of bound-state quantum electrodynamics (QED) methods and suppressed electron correlation effects. For neutral atoms QED effects are included in calculations either within the rigorous QED approach~\cite{Shabaev:02a,Labzowsky:1998,Mohr:98b,Lindgren:2004,Andreev:2008} or within the model QED Hamiltonians~\cite{Shabaev:13,Flambaum:2005,Pyykko:2003,Thierfelder:2010,Schwerdtfeger:2017,LeimbachAt:2020} (see also references therein). 

A number of studies is devoted to the treatment of QED effects in light-atom \textit{molecules} such as H$_2$~\cite{Pachucki:2011,Salam:2014} (and references therein) or heavy quasi-molecules with one electron~\cite{Artemyev:2015,Tupitsyn:2017}.
The present most sensitive experiments that are aimed to measure such fundamental properties as the electron electric dipole moment employ neutral or singly ionized heavy-atom molecules~\cite{ACME:18,Cornell:2017,Hudson:11a,kozyryev2017precision,Ruiz:2019,BaF:2018}. The accuracy of such experiments~\cite{ACME:18,Cornell:2017,Hudson:11a} has already surpassed the accuracy of atomic experiments~\cite{Regan:02}. However, up to now no theoretical predictions of transition energies and other properties of these molecules that take into account QED effects beyond the Dirac-Coulomb(-Breit) Hamiltonian have been reported. 
The availability of accurate theoretical prediction of transition energies for this class of molecules can considerably save the experimental time that is required to scan a certain wavelength range~\cite{Ruiz:2019}. One more area in which molecular QED effects are of importance is the physics and chemistry of superheavy elements (SHE). Corresponding experiments are complicated as one have to operate with extremely small concentrations, even with several atoms~\cite{Schwerdtfeger:2015}. The accurate predictions of transition energies in such molecules become even more important after the superheavy elements Factory has been launched~\cite{SHEFactory:2020}.

The radium monofluoride molecule, RaF, is considered as a perspective system that can be used to measure the electron electric dipole moment~\cite{Kudashov:14,Isaev:2010,Isaev:2012,Ruiz:2019,Petrov:2020,Sudip:2016b,Borschevsky:13}. The uncertainty of such measurement is inversely proportional to the coherence time. The key feature of this molecule is that the Franck-Condon matrix elements between the ground $X~^2\Sigma_{1/2}$ and the first excited $A~^2\Pi_{1/2}$ states are highly diagonal, allowing one to use this transition to laser-cool the molecule and increase the coherence time.  Recently, the first laser spectroscopy measurement on this radioactive molecule has been reported~\cite{Ruiz:2019}.

Below we describe a method that can be used to treat QED effects in molecules based on the model QED approach. We first apply it to calculate transition energies of low-lying states as well the ionization potential (IP) of the $^{226}$Ra$^+$ cation that agree with experiment within a few meV surpassing the accuracy of previous theoretical predictions. Then we apply the approach to predict  the electronic $X~^2\Sigma_{1/2} \to A~^2\Pi_{1/2}$ transition energy in RaF. 
The importance of considering QED effects arises when electron correlation effects are taken into account with the uncertainty smaller than the magnitude of QED effects. Such a problem itself is a challenging one. 
For considered systems, we achieve the required accuracy and explore at which stage QED effects should be included.
\section{Theory}
We consider the Dirac-Coulomb (DC) Hamiltonian as the starting approximation:
\begin{eqnarray}
 \nonumber
 H_{DC} &=& \sum\limits_j\, \left[ c\, \boldsymbol{\alpha}_j \cdot {\bf{p}}_j + \beta_j c^2 
+V_{\rm nuc}(j) \right]
+ \sum\limits_{j<k}\,
\frac{1}{r_{jk}},
 \label{EQ:HAMILTONIAN}
\end{eqnarray}
where $\alpha$, $\beta$ are Dirac matrices, $V_{\rm nuc}$ is the potential due to the nuclear subsystem of the molecule and summation is over all electrons. The main two-electron correction to this Hamiltonian is the Breit interelectronic interaction. The leading part of this interaction is the Gaunt term, which has been considered in this paper. A number of methods has been developed for many-electron atoms to go beyond the Dirac-Coulomb-Breit Hamiltonian. A very prospective approach is the model QED  Hamiltonian~\cite{Shabaev:13,Flambaum:2005,Pyykko:2003}. It consists of the vacuum-polarization (VP) and self-energy (SE) terms:
\begin{equation}
\label{HQED}
 H_{QED}=H_{VP}+H_{SE}.
\end{equation}
The model QED approach can be combined with well-developed methods to treat high-order electron correlation effects that are important for neutral heavy atoms. The dominant part of $H_{VP}$ can be represented by the Uehling potential. For the case of a finite nucleus, some approximate formulas are derived, e.g. in ~\onlinecite{Ginges:2016}. A number of expressions has been suggested for the form of the more complicated self-energy model Hamiltonian~$H_{SE}$.
Such an operator implies a scaling of the Lamb shift result for the Coulomb potential~\cite{Shabaev:13,Flambaum:2005,Pyykko:2003,Thierfelder:2010,Schwerdtfeger:2017,LeimbachAt:2020,Indelicato:1990,Tupitsyn:2013,Lowe:2013}.
It is suggested to use matrix elements for the H-like ion to predict results for the corresponding many-electron atom. In particular, this is possible due to the proportionality property of radial parts of heavy atoms wavefunctions having different principal quantum numbers $n$ and the same relativistic quantum number $\kappa=(-1)^{j+l+1/2}(j+1/2)$
\cite{Khriplovich:91,Titov:96,Titov:99,Shabaev:01a, Mosyagin:10a,Dzuba:2011,Titov:14a,Skripnikov:15b,Mosyagin:16,Skripnikov:16a,Skripnikov:2020e}. 

We are interested in the expression for the $H_{SE}$ operator, which can be conveniently used in molecular calculations providing good accuracy. Let us consider a reduced one-particle density matrix $\rho(\mathbf{r}|\mathbf{r'})$ obtained from correlation calculation of some molecule containing a heavy nucleus with the origin chosen at the position of this nucleus:
\begin{equation}
\begin{array}{l}                 
\rho(\mathbf{r}|\mathbf{r'}) = \sum\limits_{p,q} \rho_{p,q} \varphi_{p}(\mathbf{r})\varphi_{q}^{\dagger}(\mathbf{r'}),
 \label{rrofull}
\end{array}
\end{equation}
where $\{\varphi_{p}\}$ are molecular bispinors. The expectation value of some one-particle operator $X$ that is localized within a sphere with a small radius $R_c$ around the heavy nucleus can be calculated as:
\begin{equation} 
   \langle {X} \rangle = \sum_{p,q} \rho_{p, q}\int \varphi_{q}^{\dagger}X\varphi_{p}d\mathbf{r} \approx
   \sum_{p,q} \rho_{p, q}\int _{|\mathbf{r}|<R_c} \varphi_{q}^{\dagger}X\varphi_{p}d\mathbf{r}.
 \label{OperFull} 
\end{equation} 
In this region, one can reexpand molecular bispinors $\{\varphi_{p}\}$ in terms of some sufficiently complete set of functions $\eta_{kljm}(\mathbf{r})$ centered at the heavy nucleus:
\begin{equation} 
   \varphi_{p}(\mathbf{r}) \approx \sum_{kljm} C^p_{kljm} \eta_{kljm}(\mathbf{r}),~~|\mathbf{r}| \le R_c,
 \label{expans}     
\end{equation} 
where $C^p_{kljm}$ are expansion coefficients.
One can see from Eqs.~(\ref{expans}) and (\ref{OperFull})
that it is sufficient to know matrix elements $\int \eta_{nljm}^{\dagger}X\eta_{n'l'j'm'}d\mathbf{r}\ $ in order to calculate the expectation value of operator $X$.

Let us introduce functions $\widetilde{h}_{kljm}(\mathbf{r})$, defined as:
\begin{equation}
  \widetilde{h}_{kljm}(\mathbf{r})=\eta_{kljm}(\mathbf{r}) \theta(R_c-|\mathbf{r}|),
\label{hfuns}  
\end{equation}
where $\theta(R_c-|\mathbf{r}|)$ is the Heaviside step function.
If we choose orthogonal set of functions $h_{kljm}(\mathbf{r})$ that are linear combinations of $\widetilde{h}_{kljm}$ then
one can define the model $X=H_{SE}$ operator as follows:
\begin{equation}
    H_{SE} \approx \sum_{kljm, k'l'j'm'} \frac{|h_{kljm}\rangle X_{kljm,k'l'j'm'} \langle h_{k'l'j'm'}|}{\langle h_{kljm}|h_{kljm}\rangle\langle h_{k'l'j'm'}|h_{k'l'j'm'}\rangle}.
\label{Xmod}    
\end{equation}
Note, that the model SE operator~(\ref{Xmod}) is diagonal in $l,j,m$, but not in $k$. The model SE operator proposed in Ref.~\onlinecite{Shabaev:13} has a more general form than the operator~(\ref{Xmod}). In this Ref. one additionally separates a semilocal part. For the convenience of the present molecular implementation, we do not separate such a term.

\begin{table*}
\caption{Calculated ionization potential and transition energies of low-lying electronic states of the Ra$^+$ cation.
Deviations from the experimental values~\cite{AtomicSpectra:2005} 
are given in the square brackets. All values are in meV.}
\label{RaResults}
\begin{tabular*}{0.9\textwidth}
{@{\extracolsep{\fill}}lrrrrr}
\hline
\hline
 Method             & 7s $^2$S$_{1/2}$ IP  & 6d $^2$D$_{3/2}$ & 6d $^2$D$_{5/2}$& 7p $^2$P$_{1/2}$ & 7p $^2$P$_{3/2}$ \\
\hline
DHF                 & 9410.3  & 1679.1 & 1773.8 & 2358.3 & 2850.8 \\
CCSD(T)             & 10151.5 & 1540.3 & 1747.6 & 2653.2 & 3259.3 \\
High harmonics, CBS & 14.8    & -15.2  & -13.6  & 5.4    & 7.3    \\
CCSDTQ(P)-CCSD(T)   & -5.4    & -1.4   & -2.4   & -3.4   & -3.7   \\
Gaunt               & -2.0    & -11.1  & -14.1  & 5.3    & 0.4    \\
QED                 & -10.9   & -15.3  & -14.2  & -11.1  & -10.7  \\
Final               & 10148.1 [ -1.0] & ~~~~1497.3 [~~1.0] & ~~~~1703.3 [~~0.6] & ~~~~2649.3 [ -2.1] & ~~~~3252.6 [ -3.1] \\
\multicolumn{6}{c}{Other theory:}    \\ 
QED~\cite{Ginges:2015}    & -10.8    & -15.5    & -14.6    & -11.0    & -10.5    \\
Breit~\cite{Eliav:1996}   & -2.2     & -9.7    & -12.8     &  4.5    & -0.5\\
CP+SD, DCB+QED~\cite{Ginges:2015}& 10127.5 [ 19.7] & 1485.0 [ 13.3]   & 1699.9 [~~4.0]   & 2626.6 [ 20.6]   & 3232.6 [ 16.8]   \\
CP, DCB, QED~\cite{Dinh:2008}    & 10170.9 [-23.7] & ---      & ---      & 2639.7 [~~7.5]   & 3256.2 [~~-6.7]         \\
CP+SD, DCB+QED~\cite{Dzuba:2013} & 10131.2 [ 15.9] & 1478.6 [ 19.6]  & 1694.7 [~~9.2]  & 2628.8 [ 18.4] & 3234.6 [ 14.9]   \\
FS-CCSD, DCB~\cite{Eliav:1996}    & 10169.8 [-22.6] & 1541.0 [-42.7]   & 1746.8 [-42.9]  & 	2673.5 [-26.2] & 3272.7 [-23.2]    \\
All order, DCB, QED~\cite{Pal:2009}  & 10105.7 [ 41.5] & 1490.3 [~~8.0]   & 1681.3 [ 22.6]   & 2626.2 [ 21.0]   & 3222.2 [ 27.3]   \\
\\
Experiment~\cite{AtomicSpectra:2005}          & 10147.2~~~~~~~~~~ & 1498.3~~~~~~~~~   & 1703.9~~~~~~~~~   & 2647.2~~~~~~~~~   & 3249.5~~~~~~~~~    \\
\hline
\hline
\end{tabular*}
\end{table*}

In Ref.~\onlinecite{Shabaev:13} matrix elements of the self-energy over H-like functions have been calculated and tabulated for many elements of the Periodic Table, and a corresponding extrapolation procedure has been pointed out to obtain them for any element. Both diagonal and off-diagonal elements have been calculated~\cite{Shabaev:13}. The latter have been shown to be valuable for the spectral representation of the model SE operator~\cite{Dyall:2013}. In Eq.~(\ref{Xmod}) it is possible to use different choices of basis functions $h_{kljm}$.
They should form a complete (and orthogonal) set inside the corresponding sphere. In practice, a large set of such functions (\ref{hfuns}) can lead to a numerical linear dependence. To overcome this, we first consider H-like functions $\eta_{nljm}$ and construct corresponding functions $\widetilde{h}_{nljm}$ using Eq.~(\ref{hfuns}). Then we construct the overlap matrix, having the scalar products of normalized functions $\widetilde{h}_{nljm}$ as matrix elements. As the final set of functions $h_{kljm}$ for Eq.~(\ref{Xmod}) we choose eigenvectors of this matrix having eigenvalues larger than $10^{-6}$. Such a choice of functions reasonably solves the linear dependence problem and  related to the canonical orthonormalization procedure~\cite{Lowdin:1956}.
Another procedure has been used in~\cite{Shabaev:13}. 
The procedure described above allowed us to use an analytic integration in present molecular calculations.  We have used values of $R_c$ which correspond to radii of cutting functions in~\cite{Shabaev:13}.

The contribution of QED effects described by the operator~(\ref{Xmod}) can be calculated as an expectation value using the density matrix formalism or can be extracted by adding  the operator~(\ref{Xmod}) to the molecular Hamiltonian. In the latter case, one can also obtain the second and higher orders contributions to the energy of the considered system with respect to this interaction.

Dirac-Hartree-Fock and coupled cluster calculations up to the coupled cluster with single, double, and perturbative triple cluster amplitudes, CCSD(T), level were performed with the {\sc dirac}~\cite{DIRAC15} code. Higher-order coupled cluster calculations were performed with the~{\sc mrcc}~\cite{MRCC2020,Kallay:1} code. Scalar relativistic correlation calculations to ensure the basis set  completeness and to generate compact basis sets were performed using the {\sc cfour}~\cite{CFOUR} code. 
The code developed in this work has been employed to calculate the QED contribution to molecular and atomic transition energies.

\section{Results and discussion}

Table~\ref{RaResults} gives the calculated values of IP and transition energies of the low-lying electronic states of the $^{226}$Ra$^+$ cation. The largest deviation from the experimental value is found for the 7p$^2$P$_{3/2}$ state. Main calculations have been performed within the CCSD(T) method using the Dirac-Coulomb Hamiltonian and correlating all electrons. The energy cutoff for virtual orbitals included in the correlation calculation has been set to 10000 hartree~\cite{Skripnikov:17a,Skripnikov:15a}. In this calculation, the modified Gaussian-type Dyall's AEQZ~\cite{Dyall:12} basis set has been employed that consists of 42s,38p,27d,17f,11g,3h,2i functions: we have added diffuse functions of s$-$, p$-$, d$-$ and f$-$ types and replaced  g$-$, h$-$ and i$-$ by the natural basis functions, constructed as described in~\cite{Skripnikov:2020e,Skripnikov:13a}. The ``High harmonics'' contribution line in Table~\ref{RaResults} gives contribution of additional d$-$, f$-$, g$-$, h$-$, i$-$ as well as k$-$ and l$-$ type functions. The contribution of functions up to l=6 (42s,38p,35d,27f,13g,9h,6i functions) has been calculated within the relativistic CCSD(T) method, while the contribution of 6 k$-$ and 6 l$-$ type functions has been calculated within the relativistic Fock-Space coupled cluster with single and double cluster amplitudes method, FS-CCSD. $1s..3d$ electrons were not included in the correlation calculations as they gave negligible contribution in the main calculation. We have also included the extrapolated contribution of higher harmonics to treat a complete basis set limit, CBS. It was found that the ``High harmonics'' contribution can be calculated rather accurately using the valence scalar-relativistic version of the generalized relativistic effective core potential, GRECP, approach for 27 valence and outer core electrons~\cite{Titov:99,Mosyagin:10a,Mosyagin:16,Mosyagin:17} instead of the Dirac-Coulomb Hamiltonian. For the IP and all of transition energies considered for the Ra$^+$ cation, such an approach reproduces the values obtained within the DC Hamiltonian within about 1 meV. This implies a significant simplification in calculating this contribution in molecules, where the symmetry is lower.
For atomic calculation, this approach allowed us to consider the contribution of further extension of the basis set for high harmonics --- up to 25 g, 25 h, and 20 i functions, that was not practically possible within the DC Hamiltonian. The contribution of the Gaunt interelectron interaction has been treated at the FS-CCSD level within the molecular-mean-field exact 2-component approach~\cite{Sikkema:2009}. To treat higher-order correlations effects, we have used a procedure in which compact basis sets~\cite{Skripnikov:13a,Skripnikov:16b,Skripnikov:2020e} have been employed, and a smaller number of electrons were correlated. In calculations within the coupled cluster with single, double, triple, and perturbative quadruple cluster amplitudes method, CCSDT(Q), $1s..4f$ electrons of Ra have been excluded from the correlation treatment while in calculations with iterative quadruple and perturbative five-fold amplitudes within the CCSDTQ(P) method~\cite{Kallay:6}, $5s5p$ electrons have been also frozen and the two-component GRECP approach has been used. The latter allowed us to use a very compact contracted basis set~\cite{Skripnikov:13a,Skripnikov:16b,Skripnikov:2020e}. To calculate QED contribution, the developed model SE operator (\ref{Xmod}) as well as $H_{VP}$ have been added to the DC Hamiltonian before the correlation CCSD(T) stage. For comparison, the values of the transition energies and IP are given at the Dirac-Hartree-Fock (DHF) level. It is estimated that the theoretical uncertainty of the final atomic values is less than 5 meV and is dominated by the basis set imperfections.

Table~\ref{RaResults} gives also the theoretical values of the IP and transition energies of the Ra$^+$ cation calculated previously. One can see good agreement between QED contributions calculated in the present paper and in Ref.~\onlinecite{Ginges:2015}. The calculated Gaunt interaction contributions are close (with the largest deviation of 1.4 meV) to the full zero-frequency Breit interaction contributions calculated within the FS-CCSD method in Ref.~\onlinecite{Eliav:1996}.  

\begin{figure}[]
\centering
\includegraphics[width = 3.0 in]{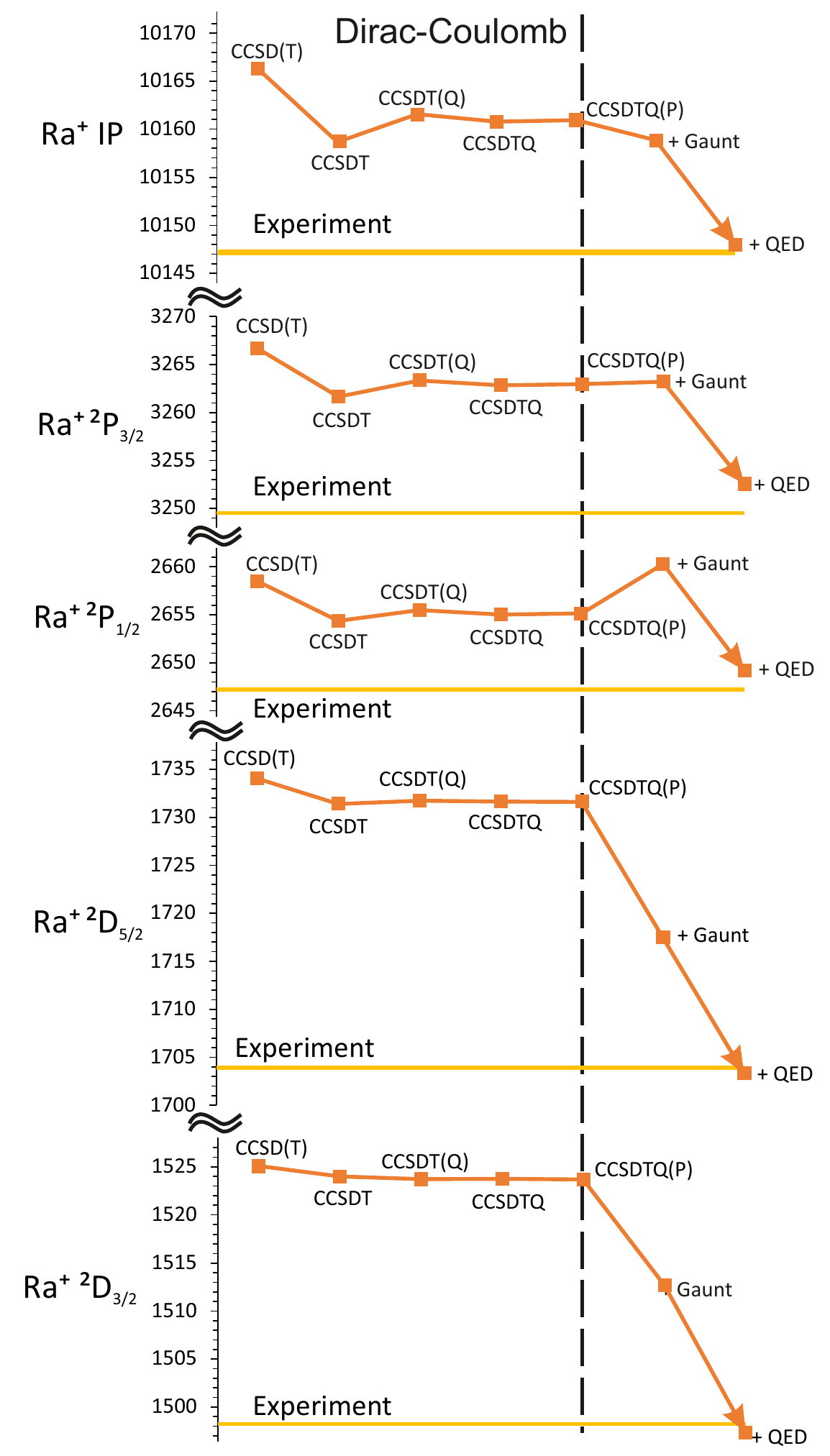}
 \caption{Contributions of electron correlation effects, Gaunt and QED effects into transition energies of low-lying states of Ra$^+$ and its ionization potential. All values are in meV.}
 \label{FigRa}
\end{figure}

Figure~\ref{FigRa} illustrates contributions of correlation effects at different levels of theory, Gaunt and QED effects for the IP and transition energies of the low-lying electronic states of Ra$^+$.  It can be seen that for all cases, QED contribution is larger than the correlation contribution beyond the CCSD(T) level as well as the Gaunt contribution, except for 6d $^2D_{5/2}$ and 6d $^2D_{3/2}$ states, where the Gaunt and QED contribution have similar magnitudes.

Table~\ref{RaFResults} gives the value of the transition energy between the ground X$^2\Sigma_{1/2}$ and the first excited state A$^2\Pi_{1/2}$ of the RaF molecule, which can be used to laser-cool this molecule~\cite{Ruiz:2019}. For both electronic states the internuclear distance of 4.23 Bohr~\cite{Skripnikov:2020e,Kudashov:14} has been used. In the main CCSD(T) calculation, the same basis set on Ra as in atomic calculation has been used. For F, the AETZ~\cite{Dyall:12} basis set has been used. All 97 electrons of RaF have been correlated with virtual energy cutoff equal to 10000 hartree as in atomic calculation. To consider the basis set extension contribution, we have increased the basis set on F up to AAEQZ~\cite{Dyall:12} and basis set on Ra up to 42s,38p,27d,27f,13g,9h,6i functions. This calculation has been performed within the FS-CCSD method with excluded $1s..3d$ electrons of Ra. In the ``High harmonics'' contribution we have also included contributions of additional d-, g-, h- and i-type function on Ra within the scalar-relativistic approach, which has been 
justified in atomic calculations above. We have also included the extrapolated contribution of even higher harmonics on Ra as in the atomic case. Interestingly, that the ``High harmonics'' contribution given in Table~\ref{RaFResults} for the molecular case was found to be much smaller than in the atomic case. This can be partly explained by the effective enrichment of the Ra basis set in the molecule due to functions centered on F. Contribution of iterative triple and perturbative quadruple cluster amplitudes has been obtained within the CCSDT(Q) method. Outer 27 electrons of RaF has been included in this correlation calculation and an optimized compact basis set~\cite{Skripnikov:13a,Skripnikov:16b,Skripnikov:2020e} has been used.

The agreement between the final result and experiment~\cite{Ruiz:2019} is just about 2 meV. We have found only one previous theoretical prediction of the transition energy obtained within the DC Hamiltonian at the FS-CCSD level with correlation of 17 outer electrons~\cite{Isaev:2013}: depending on the basis set choice, the predicted value is 1737 meV or 1644 meV.

\begin{table}[]
\caption{Calculated $X ^2\Sigma_{1/2} \to A ^2\Pi_{1/2}$  transition energy T$_{\rm e}$ in the RaF molecule in meV. Deviation from the experimental value~\cite{Ruiz:2019}
is given in the square brackets. }
\label{RaFResults}
\begin{tabular}{lr}
\hline
\hline
Method                     & T$_{\rm e}$, meV \\
\hline          
CCSD(T)                    & 1659.0 \\
High harmonics, CBS        & 0.2    \\
CCSDT(Q)-CCSD(T)           & -3.4   \\
Total, Dirac-Coulomb       & 1655.8 \\
Gaunt                      & 0.6    \\
QED                        & -7.0   \\
Final                      & 1649.4 [-1.9] \\
Experiment~\cite{Ruiz:2019}   & 1647.5 \\
\hline
\hline
\end{tabular}
\end{table}

Qualitatively, both considered electronic states of RaF have one unpaired electron on the so-called non-bonding atomic-like orbitals. However, the present direct treatment of the QED effects in RaF shows that for the considered electronic transition, QED contribution is not equal to that for the corresponding transition in Ra$^+$.

\begin{figure}[]
\centering
\includegraphics[width = 3.0 in]{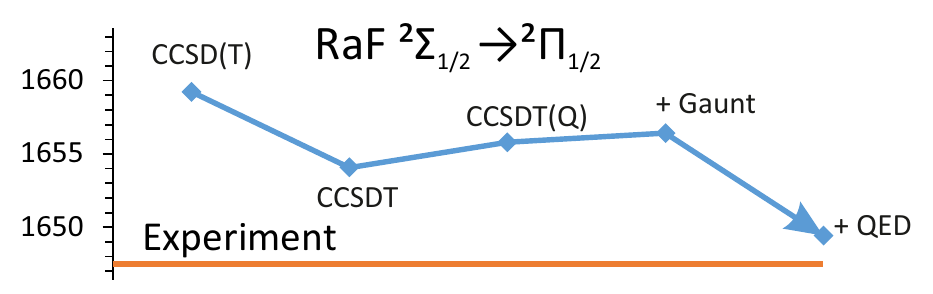}
 \caption{Contributions of electron correlation effects, Gaunt and QED effects into  into the $X ^2\Sigma_{1/2} \to A ^2\Pi_{1/2}$ transition energy in meV.}
 \label{FigRaF}
\end{figure}

Figure~\ref{FigRaF} shows contributions of correlation effects at different levels of theory, Gaunt and QED effects for the transition energy in RaF. It can be seen that the QED contribution is about an order of magnitude larger than the Gaunt contribution and larger than all correlation effects beyond the CCSD(T) level as in the atomic case above. 

\section{Conclusions}

The model QED operator which can be directly used for predicting QED contributions to electronic energies of molecular states is implemented and applied to the RaF molecule and Ra$^+$ cation. 
It is shown that in both cases it was QED effects that were a missing ingredient for removing the main discrepancy between the theory and experiment even after the inclusion of high-order terms into the wavefunction expansion. Therefore, the introduction of the QED effects seems to be an important step towards the predictive molecular electronic structure theory.
This conclusion seems us to be reasonable for molecules containing SHE where QED effects are expected to be even higher and for which accurate theoretical predictions are of particular importance for preparation of experiments.

\begin{acknowledgments}
I am grateful to A.V. Titov and V.M. Shabaev for useful discussions. Electronic structure calculations have been carried out using computing resources of the federal collective usage center Complex for Simulation and Data Processing for Mega-science Facilities at NRC “Kurchatov Institute”, http://ckp.nrcki.ru/, and computers of Quantum Chemistry Lab at NRC ``Kurchatov Institute" - PNPI.

The reported study was funded by RFBR according to the research project No. 20-32-70177.
\end{acknowledgments}

\end{document}